\documentclass[conference,10pt]{IEEEtran}
\normalsize
\ifCLASSINFOpdf
\else
\fi

\usepackage{amsthm}
\usepackage{amsmath}
\usepackage{amssymb}
\usepackage{graphicx}
\usepackage{esint}
\usepackage{amsfonts}
\usepackage{cite}
\usepackage{balance}
\usepackage{caption}
\usepackage{subcaption}
\usepackage{epstopdf}
\usepackage{color}
\usepackage{tikz}
\usepackage{pgfplots} 
\usetikzlibrary{patterns}

\usepackage[left=1.62cm,right=1.62cm,top=1.9cm]{geometry}

\makeatletter

\theoremstyle{plain}

\makeatother

\providecommand{\theoremname}{Proposition}

\begin{document}
\title{Can graph neural network-based detection mitigate the impact of hardware imperfections?}

\author{\IEEEauthorblockN{Lamprini Mitsiou\IEEEauthorrefmark{1},  Stylianos Trevlakis\IEEEauthorrefmark{1}, Argiris Tsiolas\IEEEauthorrefmark{2}, Dimitrios J. Vergados\IEEEauthorrefmark{2}, Angelos Michalas\IEEEauthorrefmark{2},  and \\ Alexandros-Apostolos A. Boulogeorgos\IEEEauthorrefmark{2}
}
\IEEEauthorrefmark{1}{{\footnotesize{}{}{{{{{{{Research \& Development Department, InnoCube P.C., Thessaloniki 55535, Greece.
E-mails: mitsiou@innocube.org, trevlakis@innocube.org.
}}}}}}}}}
\\
\IEEEauthorrefmark{2}{\footnotesize{}{}{{{{{{{Department of Electrical and Computer Engineering, University of Western Macedonia, Kozani 50100, Greece. 
}}}}}}}}\\
\footnotesize{}{}{E-mail: atsiolas@gmail.com, dvergados@uowm.gr, amichalas@uowm.gr, aboulogeorgos@uowm.gr.}
}
\maketitle	

\begin{abstract}
    Until recently, researchers used machine learning  methods to compensate for hardware imperfections at the symbol level, indicating that optimum radio-frequency transceiver performance is possible. Nevertheless, such approaches neglect the error correcting codes used in wireless networks, which inspires machine learning (ML)-approaches that learn and minimise hardware imperfections at the bit level. In the present work, we evaluate a graph neural network (GNN)-based intelligent detector's in-phase and quadrature imbalance (IQI) mitigation capabilities. We focus on a high-frequency, high-directional wireless system where IQI affects both the transmitter (TX) and the receiver (RX). The TX uses a GNN-based decoder, whilst the RX uses a linear error correcting algorithm. The bit error rate (BER) is computed using appropriate Monte Carlo simulations to quantify performance. Finally, the outcomes are compared to both traditional systems using conventional detectors and wireless systems using belief propagation based detectors. Due to the utilization of graph neural networks, the proposed algorithm is highly scalable with few training parameters and is able to adapt to various code parameters. 
\end{abstract}
\begin{IEEEkeywords}
Belief propagation, bit error rate, graph neural networks, hardware imperfection mitigation, in-phase and quadrature imbalance, machine learning.
\end{IEEEkeywords}

\section{Introduction}\label{S:Intro}

As the wireless world search for unexploited resources in higher frequency bands, like millimeter wave and terahertz, new challenges are identifies and call for novel solutions~\cite{Boulogeorgos2018,A:Analytical_Performance_Assessment_of_THz_Wireless_Systems,Boulogeorgos2019,9615497,trevlakis2023localization}. One of the most important challenges is dealing with the impact of transceiver hardware imperfections. As discussed in~\cite{A:Effects_of_RF_impairments_In_Cascaded,9039743,8094903,7460238,9159653,7496991,9145251}, hardware imperfections, such as local oscillators' phase noise, amplifier's non-linearity and especially up and down-converters in-phase and quadrature imbalance (IQI), significantly limit the reliability of high-frequency wireless systems. Note that, as described~\cite{8933807}, the hardware imperfection of wireless systems in the higher frequency range cannot be completely avoided, even with new technological solutions supported by integrated microwave photonics.  

Motivated by this, several researchers have presented hardware imperfections mitigation solutions~\cite{A:Frequency_selective_IQ_mismatch_callibration_of_wideband_DC_transmitters,8527635,8383719,9605580,8476222}. In particular, in~\cite{A:Frequency_selective_IQ_mismatch_callibration_of_wideband_DC_transmitters}, the authors reported a widely linear IQI calibration structure that estimates the IQI parameters using either second-order statistics or least-square-based model fitting. The authors of~\cite{8527635} used higher order statistics-based approaches in order to estimate the amplifier non-linearity in the presence of IQI and documented an IQI parameter maximum-likelihood estimation approach. The aforementioned approaches are two-step processes that are usually energy consuming. 

To counterbalance this, the authors of~\cite{8383719} presented a real-valued  time-delay neural network that is used as an one-step mitigation process; thus, simplifies the compensation process. In~\cite{9605580}, a shortcut real-valued time-delay neural network for compensating IQI and amplifiers non-linearity was introduced. Finally, in~\cite{8476222}, a neural network-based digital predistortion as a solution to countermeasure the impact of cross-talk, amplifier non-linearity, IQI, and direct current offset was~presented. 

In other words, the authors of~\cite{8527635,9605580,8476222} aimed to employ machine learning methodologies in order to compensate the impact of hardware imperfections at a symbol-level, proving that with such approaches the ideal radio-frequency (RF) transceiver performance are reachable. However, following such approach, it is impossible to exploit the characteristics of the error correction codes that are employed in nowadays wireless systems. This observation motivates the design of machine learning (ML)-approaches that learn and de-emphasize the impact of hardware imperfections in a bit level. These approaches should be scalable and have a relatively low-number of training parameters in order to be adaptive in different code parameters. Motivated by this as well as the close relation between the Tanner graphs, which can be used to represent codes, and the graph neural networks (GNNs), in this paper, we assess the IQI mitigation capabilities of an intelligent detector that employs GNN. In particular, we consider a high-frequency and high-directional wireless system in which both the transmitter (TX) and the receiver (RX) suffer from IQI. A linear error correction code is employed by the TX, while a GNN-based decoder is used by the RX. To quantify the performance, the bit error rate (BER) is derived through respective Monte Carlo simulations. The results are benchmarked against conventional systems that employ traditional detectors, as well as wireless systems that use belief propagation (BF) based detectors.   

The organization of the rest of the paper is as follows: The system model is described in Section~\ref{sec:SSM}. Section~\ref{SS:Train} reports the operation and training procedures of the intelligent detectors. Results and related discussions are documented in Section~\ref{S:Results}. Finally, the conclusions and main message of this contribution is summarized in Section~\ref{S:Con}.  

\subsubsection*{Notations} In what follows, $\left[\cdot|| \cdot\right]$ stands for the concatenation operator. Moreover, $\oplus$ represents the message aggregation function, which, for this contribution is the mean value. The logarithm to base $2$ is denoted as $\log_2\left(\cdot\right)$. The sum of $x_i$ for $i\in[1, N]$ is represented as $\sum_{i=1}^{N}x_i$, while $\prod_{i=1}^{N}x_i$ is the product of $x_i$ for $i\in[1,N]$. Finally, $\Pr\left(\mathcal{E}\right)$ is the probability of the event $\mathcal{E}$.  

\section{System model }\label{sec:SSM}

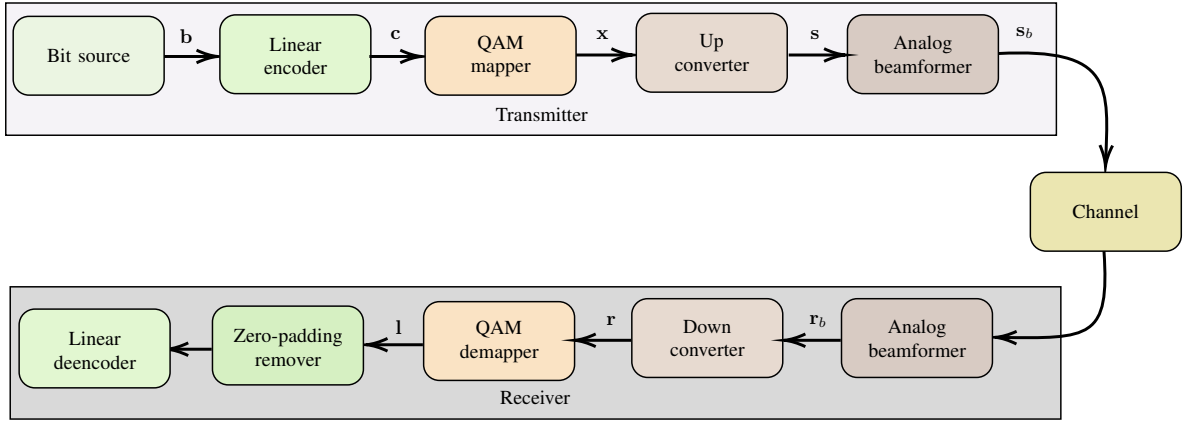
\begin{figure*}
	\centering
	\scalebox{0.75}{

		\tikzset{every picture/.style={line width=0.75pt}} 
		
		\begin{tikzpicture}[x=0.75pt,y=0.75pt,yscale=-1,xscale=1]
			
			\draw  [fill={rgb, 255:red, 236; green, 232; blue, 240 }  ,fill opacity=0.5 ] (3,5) -- (709.5,5) -- (709.5,94) -- (3,94) -- cycle ;
			\draw  [fill={rgb, 255:red, 235; green, 245; blue, 226 }  ,fill opacity=1 ] (8,24.6) .. controls (8,18.75) and (12.75,14) .. (18.6,14) -- (98.9,14) .. controls (104.75,14) and (109.5,18.75) .. (109.5,24.6) -- (109.5,56.4) .. controls (109.5,62.25) and (104.75,67) .. (98.9,67) -- (18.6,67) .. controls (12.75,67) and (8,62.25) .. (8,56.4) -- cycle ;
			\draw [line width=1.5]    (109.5,41) -- (142.5,41) ;
			\draw [shift={(145.5,41)}, rotate = 180] [color={rgb, 255:red, 0; green, 0; blue, 0 }  ][line width=1.5]    (14.21,-4.28) .. controls (9.04,-1.82) and (4.3,-0.39) .. (0,0) .. controls (4.3,0.39) and (9.04,1.82) .. (14.21,4.28)   ;
			\draw  [fill={rgb, 255:red, 226; green, 247; blue, 209 }  ,fill opacity=1 ] (147,23.6) .. controls (147,17.75) and (151.75,13) .. (157.6,13) -- (237.9,13) .. controls (243.75,13) and (248.5,17.75) .. (248.5,23.6) -- (248.5,55.4) .. controls (248.5,61.25) and (243.75,66) .. (237.9,66) -- (157.6,66) .. controls (151.75,66) and (147,61.25) .. (147,55.4) -- cycle ;
			\draw [line width=1.5]    (247.5,41) -- (279.5,41) ;
			\draw [shift={(282.5,41)}, rotate = 180] [color={rgb, 255:red, 0; green, 0; blue, 0 }  ][line width=1.5]    (14.21,-4.28) .. controls (9.04,-1.82) and (4.3,-0.39) .. (0,0) .. controls (4.3,0.39) and (9.04,1.82) .. (14.21,4.28)   ;
			\draw  [fill={rgb, 255:red, 250; green, 227; blue, 196 }  ,fill opacity=1 ] (285,24.6) .. controls (285,18.75) and (289.75,14) .. (295.6,14) -- (375.9,14) .. controls (381.75,14) and (386.5,18.75) .. (386.5,24.6) -- (386.5,56.4) .. controls (386.5,62.25) and (381.75,67) .. (375.9,67) -- (295.6,67) .. controls (289.75,67) and (285,62.25) .. (285,56.4) -- cycle ;
			\draw [line width=1.5]    (386.5,40) -- (421.5,40) ;
			\draw [shift={(424.5,40)}, rotate = 180] [color={rgb, 255:red, 0; green, 0; blue, 0 }  ][line width=1.5]    (14.21,-4.28) .. controls (9.04,-1.82) and (4.3,-0.39) .. (0,0) .. controls (4.3,0.39) and (9.04,1.82) .. (14.21,4.28)   ;
			\draw  [fill={rgb, 255:red, 231; green, 218; blue, 208 }  ,fill opacity=1 ] (427,22.6) .. controls (427,16.75) and (431.75,12) .. (437.6,12) -- (517.9,12) .. controls (523.75,12) and (528.5,16.75) .. (528.5,22.6) -- (528.5,54.4) .. controls (528.5,60.25) and (523.75,65) .. (517.9,65) -- (437.6,65) .. controls (431.75,65) and (427,60.25) .. (427,54.4) -- cycle ;
			\draw  [fill={rgb, 255:red, 216; green, 203; blue, 195 }  ,fill opacity=1 ] (569,23.6) .. controls (569,17.75) and (573.75,13) .. (579.6,13) -- (659.9,13) .. controls (665.75,13) and (670.5,17.75) .. (670.5,23.6) -- (670.5,55.4) .. controls (670.5,61.25) and (665.75,66) .. (659.9,66) -- (579.6,66) .. controls (573.75,66) and (569,61.25) .. (569,55.4) -- cycle ;
			\draw [line width=1.5]    (529.5,40) -- (564.5,40) ;
			\draw [shift={(567.5,40)}, rotate = 180] [color={rgb, 255:red, 0; green, 0; blue, 0 }  ][line width=1.5]    (14.21,-4.28) .. controls (9.04,-1.82) and (4.3,-0.39) .. (0,0) .. controls (4.3,0.39) and (9.04,1.82) .. (14.21,4.28)   ;
			\draw [line width=1.5]    (670.5,39) .. controls (748.32,35.06) and (744.63,61.2) .. (742.59,112.63) ;
			\draw [shift={(742.5,115)}, rotate = 272.16] [color={rgb, 255:red, 0; green, 0; blue, 0 }  ][line width=1.5]    (14.21,-4.28) .. controls (9.04,-1.82) and (4.3,-0.39) .. (0,0) .. controls (4.3,0.39) and (9.04,1.82) .. (14.21,4.28)   ;
			\draw  [fill={rgb, 255:red, 235; green, 230; blue, 177 }  ,fill opacity=1 ] (692,129.6) .. controls (692,123.75) and (696.75,119) .. (702.6,119) -- (782.9,119) .. controls (788.75,119) and (793.5,123.75) .. (793.5,129.6) -- (793.5,161.4) .. controls (793.5,167.25) and (788.75,172) .. (782.9,172) -- (702.6,172) .. controls (696.75,172) and (692,167.25) .. (692,161.4) -- cycle ;
			\draw  [fill={rgb, 255:red, 180; green, 180; blue, 180 }  ,fill opacity=0.5 ] (6,196) -- (712.5,196) -- (712.5,285) -- (6,285) -- cycle ;
			\draw [line width=1.5]    (741.5,172) .. controls (745.44,231.1) and (732.89,231.02) .. (671.35,230.04) ;
			\draw [shift={(668.5,230)}, rotate = 0.9] [color={rgb, 255:red, 0; green, 0; blue, 0 }  ][line width=1.5]    (14.21,-4.28) .. controls (9.04,-1.82) and (4.3,-0.39) .. (0,0) .. controls (4.3,0.39) and (9.04,1.82) .. (14.21,4.28)   ;
			\draw  [fill={rgb, 255:red, 216; green, 203; blue, 195 }  ,fill opacity=1 ] (565,214.6) .. controls (565,208.75) and (569.75,204) .. (575.6,204) -- (655.9,204) .. controls (661.75,204) and (666.5,208.75) .. (666.5,214.6) -- (666.5,246.4) .. controls (666.5,252.25) and (661.75,257) .. (655.9,257) -- (575.6,257) .. controls (569.75,257) and (565,252.25) .. (565,246.4) -- cycle ;
			\draw  [fill={rgb, 255:red, 231; green, 218; blue, 208 }  ,fill opacity=1 ] (424,214.6) .. controls (424,208.75) and (428.75,204) .. (434.6,204) -- (514.9,204) .. controls (520.75,204) and (525.5,208.75) .. (525.5,214.6) -- (525.5,246.4) .. controls (525.5,252.25) and (520.75,257) .. (514.9,257) -- (434.6,257) .. controls (428.75,257) and (424,252.25) .. (424,246.4) -- cycle ;
			\draw [line width=1.5]    (529.5,232) -- (564.5,232) ;
			\draw [shift={(526.5,232)}, rotate = 0] [color={rgb, 255:red, 0; green, 0; blue, 0 }  ][line width=1.5]    (14.21,-4.28) .. controls (9.04,-1.82) and (4.3,-0.39) .. (0,0) .. controls (4.3,0.39) and (9.04,1.82) .. (14.21,4.28)   ;
			\draw  [fill={rgb, 255:red, 250; green, 227; blue, 196 }  ,fill opacity=1 ] (284,216.6) .. controls (284,210.75) and (288.75,206) .. (294.6,206) -- (374.9,206) .. controls (380.75,206) and (385.5,210.75) .. (385.5,216.6) -- (385.5,248.4) .. controls (385.5,254.25) and (380.75,259) .. (374.9,259) -- (294.6,259) .. controls (288.75,259) and (284,254.25) .. (284,248.4) -- cycle ;
			\draw [line width=1.5]    (388.5,232) -- (423.5,232) ;
			\draw [shift={(385.5,232)}, rotate = 0] [color={rgb, 255:red, 0; green, 0; blue, 0 }  ][line width=1.5]    (14.21,-4.28) .. controls (9.04,-1.82) and (4.3,-0.39) .. (0,0) .. controls (4.3,0.39) and (9.04,1.82) .. (14.21,4.28)   ;
			\draw [line width=1.5]    (248.5,235) -- (283.5,235) ;
			\draw [shift={(245.5,235)}, rotate = 0] [color={rgb, 255:red, 0; green, 0; blue, 0 }  ][line width=1.5]    (14.21,-4.28) .. controls (9.04,-1.82) and (4.3,-0.39) .. (0,0) .. controls (4.3,0.39) and (9.04,1.82) .. (14.21,4.28)   ;
			\draw  [fill={rgb, 255:red, 214; green, 240; blue, 194 }  ,fill opacity=1 ] (142,219.6) .. controls (142,213.75) and (146.75,209) .. (152.6,209) -- (232.9,209) .. controls (238.75,209) and (243.5,213.75) .. (243.5,219.6) -- (243.5,251.4) .. controls (243.5,257.25) and (238.75,262) .. (232.9,262) -- (152.6,262) .. controls (146.75,262) and (142,257.25) .. (142,251.4) -- cycle ;
			\draw [line width=1.5]    (114.5,238) -- (141.5,238) ;
			\draw [shift={(111.5,238)}, rotate = 0] [color={rgb, 255:red, 0; green, 0; blue, 0 }  ][line width=1.5]    (14.21,-4.28) .. controls (9.04,-1.82) and (4.3,-0.39) .. (0,0) .. controls (4.3,0.39) and (9.04,1.82) .. (14.21,4.28)   ;
			\draw  [fill={rgb, 255:red, 226; green, 247; blue, 209 }  ,fill opacity=1 ] (12,221.6) .. controls (12,215.75) and (16.75,211) .. (22.6,211) -- (102.9,211) .. controls (108.75,211) and (113.5,215.75) .. (113.5,221.6) -- (113.5,253.4) .. controls (113.5,259.25) and (108.75,264) .. (102.9,264) -- (22.6,264) .. controls (16.75,264) and (12,259.25) .. (12,253.4) -- cycle ;
			
			\draw (58.75,40.5) node   [align=left] {Bit source};
			\draw (124.22,27) node    {$\mathbf{b}$};
			\draw (197.75,40.5) node   [align=left] {\begin{minipage}[lt]{39.59pt}\setlength\topsep{0pt}
					\begin{center}
						Linear \\encoder
					\end{center}
					
			\end{minipage}};
			\draw (265.22,26) node    {$\mathbf{c}$};
			\draw (335.75,40.5) node   [align=left] {\begin{minipage}[lt]{37.3pt}\setlength\topsep{0pt}
					\begin{center}
						QAM\\mapper
					\end{center}
					
			\end{minipage}};
			\draw (404.22,26) node    {$\mathbf{x}$};
			\draw (477.75,38.5) node   [align=left] {\begin{minipage}[lt]{45.24pt}\setlength\topsep{0pt}
					\begin{center}
						Up\\converter
					\end{center}
					
			\end{minipage}};
			\draw (619.75,39.5) node   [align=left] {\begin{minipage}[lt]{57.7pt}\setlength\topsep{0pt}
					\begin{center}
						Analog\\beamformer
					\end{center}
					
			\end{minipage}};
			\draw (547.22,26) node    {$\mathbf{s}$};
			\draw (688.22,25) node    {$\mathbf{s}_{b}$};
			\draw (331,74) node [anchor=north west][inner sep=0.75pt]   [align=left] {Transmitter};
			\draw (742.75,145.5) node   [align=left] {\begin{minipage}[lt]{40.73pt}\setlength\topsep{0pt}
					\begin{center}
						Channel
					\end{center}
					
			\end{minipage}};
			\draw (334,264) node [anchor=north west][inner sep=0.75pt]   [align=left] {Receiver};
			\draw (615.75,230.5) node   [align=left] {\begin{minipage}[lt]{57.7pt}\setlength\topsep{0pt}
					\begin{center}
						Analog\\beamformer
					\end{center}
					
			\end{minipage}};
			\draw (410.22,219) node    {$\mathbf{r}$};
			\draw (474.75,230.5) node   [align=left] {\begin{minipage}[lt]{45.24pt}\setlength\topsep{0pt}
					\begin{center}
						Down\\converter
					\end{center}
					
			\end{minipage}};
			\draw (549.22,219) node    {$\mathbf{r}_{b}$};
			\draw (334.75,232.5) node   [align=left] {\begin{minipage}[lt]{48.65pt}\setlength\topsep{0pt}
					\begin{center}
						QAM\\demapper
					\end{center}
					
			\end{minipage}};
			\draw (192.75,236.5) node   [align=left] {\begin{minipage}[lt]{63.4pt}\setlength\topsep{0pt}
					\begin{center}
						Zero-padding\\remover
					\end{center}
					
			\end{minipage}};
			\draw (267.22,223) node    {$\mathbf{l}$};
			\draw (62.75,238.5) node   [align=left] {\begin{minipage}[lt]{50.94pt}\setlength\topsep{0pt}
					\begin{center}
						Linear \\deencoder
					\end{center}
					
			\end{minipage}};

		\end{tikzpicture}
	}
	\caption{System model.}
	\label{Fig:SM}
\end{figure*}

As illustrated in Fig.~\ref{Fig:SM}, we consider a high-directional wireless system that consists of a TX and a RX. Both the TX and RX employ analog beamforming and their beams are assumed to be perfectly aligned.

The TX consists of a bit source that outputs a tuple of $\mathbf{b}$ bits, i.e., the codeword, which is the input of a $(N, K, L)$ linear encoder, where $N$ is the codeword length, while $K$ and $L$ are respectively the number of ones in each column and row of the parity check matrix, $\mathbf{P}$.  The linear encoder uses zero-padding in order to be able to support odd codeword lengths and outputs a bit tuple $\mathbf{c}$. 
Let $\mathcal{L}$ be the function that describes the operation of the linear encoder, then
\begin{align}
    \mathbf{c}=\mathcal{L}\{\mathbf{b}\}.
\end{align}

The output of the linear encoder is in turn inputted in a quadrature amplitude modulation (QAM) mapper. Let $\mathcal{M}\{\cdot\}$ be the function that models the operation of the QAM mapper. Then, the output of the QAM mapper can be described~as
\begin{align}
    \mathbf{x} = \mathcal{M}\{\mathbf{c}\}. 
\end{align}

The symbol vector, $\mathbf{x}$, is forwarded to the up-converter. We assume that the up-converter suffers from in-phase and quadrature imbalance. As a consequence, the baseband equivalent signal at the up-converter's output can be expressed~as~\cite{7463533} 
\begin{align}
    \mathbf{s} = K_{1}^{t}\,\mathbf{x} + K_{2}^{t}\,\mathbf{x}^{*},
    \label{Eq:s}
\end{align}
where $K_{1}^{t}$ and  $K_{2}^{t}$ are the IQI coefficients that, based on~\cite{B:Schenk-book}, can be written~as~\cite{PhD:Boulogeorgos} 
\begin{align}
    K_1^{t} = \frac{1 + g_t \, \exp\left(j\,\theta_t\right)}{2}
\end{align}
and
\begin{align}
    K_2^{t} = \frac{1 + g_t \, \exp\left(j\,\theta_t\right)}{2},
\end{align}
with $g_t$ and $\theta_t$ denoting the IQI-infused amplitude and phase mismatched, respectively. Notice that 
\begin{align}
    K_{1}^{t} = 1-\left(K_{2}^{t}\right)^{*}.
\end{align}
Moreover, the TX image rejection ratio (IRR) can be obtained~as
\begin{align}
    I_{t} = \frac{\left|K_1^t\right|^2}{\left|K_2^t\right|^2}. 
\end{align}

The up-converter is followed by the analog beamformer. The baseband equivalent at the output of the TX can be written~as
\begin{align}
    \mathbf{s}_b = \mathbf{u}\,\mathbf{s},
    \label{Eq:s_b}
\end{align}
where $\mathbf{u}$ stands for the TX beamforming vector. 

The baseband equivalent signal at the output of the RX beamformer can be expressed~as
\begin{align}
    \mathbf{r}_b = \mathbf{v}\,\mathbf{H}\,\mathbf{s}_b + \mathbf{n},
    \label{Eq:r_b}
\end{align}
where $\mathbf{v}$ and $\mathbf{H}$ stands for the RX beam-vector and the channel matrix respectively, while $\mathbf{n}$ is an additive white Gaussian noise vector. Each element of  $\mathbf{n}$ is modeled as an zero-mean complex Gaussian process of variance $N_o$. Additionally, $\mathbb{E}[n_i\,n_j]=0$, for $i\neq j$. 

With the aid of~\eqref{Eq:s_b},~\eqref{Eq:r_b} can be rewritten~as  
\begin{align}
    \mathbf{r}_b = \mathbf{v}\,\mathbf{H}\,\mathbf{u}\,\mathbf{s} + \mathbf{n}.
    \label{Eq:r_b_s2}
\end{align}
As reported in~\cite{8877185}, since the TX and RX beams are perfectly aligned, \begin{align}\mathbf{v}\,\mathbf{H}\,\mathbf{u}=h,\end{align} where $h$ is a scalar that represents the channel coefficient. As a consequence,~\eqref{Eq:r_b_s2} yields
\begin{align}
    \mathbf{r}_b = h\,\mathbf{s} + \mathbf{n}.
    \label{Eq:r_b_s3}
\end{align}
Notice that the impact of multi-path fading is respectively low. Thus, the channel coefficient models only the deterministic~path-gain. 

The output of the RX beamformer is connected to a down-converter that suffers from IQI. The down-converter introduces IQI; as a result, the baseband equivalent signal at the output of the down-converter can be expressed~as~\cite{B:Schenk-book}
\begin{align}
    \mathbf{r} = K_{1}^{r}\,\mathbf{r}_b + K_{2}^{r}\,\mathbf{r}_b^{*},
    \label{Eq:r}
\end{align}
where 
\begin{align}
    K_{1}^{r} = \frac{1+g_r\,\exp\left(-j\,\theta_r\right)}{2}
    \label{Eq:K1_r}
\end{align}
and 
\begin{align}
    K_{2}^{r} = \frac{1-g_r\,\exp\left(j\,\theta_r\right)}{2}.
    \label{Eq:K2_r}
\end{align}
In~\eqref{Eq:K1_r} and~\eqref{Eq:K2_r}, $g_r$ and $\theta_r$ are respectively the RX amplitude and phase mismatches.  The RX IRR can be written~as
\begin{align}
    \mathcal{I}_r = \frac{\left|K_{1}^{r}\right|^2}{\left|K_{2}^{r}\right|^2}. 
\end{align}
From~\eqref{Eq:r_b_s3},~\eqref{Eq:r} can be expressed~as
\begin{align}
    \mathbf{r} =K_{1}^{r}\,\left(h\,\mathbf{s} + \mathbf{n}\right) + K_{2}^{r}\,\left(h\,\mathbf{s} + \mathbf{n}\right)^{*}
\end{align}
or
\begin{align}
    \mathbf{r} =K_{1}^{r}\, h\,\mathbf{s} + K_{2}^{r}\,h\,\mathbf{s}^{*} +  K_{1}^{r}\,\mathbf{n} + K_2^{r} \mathbf{n}^{*}.
    \label{Eq:r_s2}
\end{align}
By applying~\eqref{Eq:s} to~\eqref{Eq:r_s2}, the baseband equivalent received signal at the output of the down-converter can be expressed~as
\begin{align}
    \mathbf{r} &=K_{1}^{r}\, h\,\left(K_{1}^{t}\,\mathbf{x} + K_{2}^{t}\,\mathbf{x}^{*}\right) + K_{2}^{r}\,h\,\left(K_{1}^{t}\,\mathbf{x} + K_{2}^{t}\,\mathbf{x}^{*}\right)^{*} 
    \nonumber \\ & 
    +  K_{1}^{r}\,\mathbf{n} + K_2^{r} \mathbf{n}^{*},
\end{align}
or equivalently
\begin{align}
    \mathbf{r} &=\left(K_1^{r}\, K_{1}^{t} + K_{2}^{r}\,K_2^{t} \right)\, h\,\mathbf{x} + \left(K_{1}^{r}\,K_{2}^{t} + K_{2}^{r}\,K_{1}^{t}\right)\,h\,\mathbf{x}^{*}
    \nonumber \\ & 
    +  K_{1}^{r}\,\mathbf{n} + K_2^{r} \mathbf{n}^{*}.
\end{align}
Thus, the received signal-to-distortion-plus-noise-ratio (SDNR) is given~by
\begin{align}
    \gamma = \frac{\left|K_1^{r}\, K_{1}^{t} + K_{2}^{r}\,K_2^{t}\right|^2\,\rho}{\left|K_{1}^{r}\,K_{2}^{t} + K_{2}^{r}\,K_{1}^{t}\right|^2\,\rho +  \left|K_{1}^{r}\right|^2 + \left|K_2^{r}\right|^2},
\end{align}
where $\rho$ stands for the signal-to-noise-ratio (SNR) of the ideal wireless system, i.e., the one that does not suffer from IQI, and can be expressed~as
\begin{align}
    \rho=\frac{h^2\,P_x}{N_o}.
    \label{Eq:rho}
\end{align}
In~\eqref{Eq:rho}, $P_x$ stands for the average transmission power. 

The output of the down-converter is inserted to the QAM demapper, which returns the log-likelihood ratios (LLRs) of the received signals. Let $\mathcal{D}\{\cdot\}$ represent the QAM demapper's operation. Then, the LLRs at the output of the QAM demapper can be written~as 
\begin{align}
    \mathbf{l}=\mathcal{D}\{\mathbf{r}\},
\end{align}
where the $k-$th value of $\mathbf{l}$ can be obtained~as
\begin{align}
    l_{k,m} = \log_{2}\frac{ \Pr\left(c_k=1\left|r_m\right.\right)}{\Pr\left(c_k=0\left|r_m\right.\right)}.
\end{align}
Note that $r_m$ stands for the $m-$th value of $\mathbf{m}$ that carriers the $c_k$ bit. 

The LLR vector is inputted in the zero-padding remover that outputs only the LLR elements of that corresponds to the coded message. In turn, the output of the zero-pading remover is inserted in the decoder that provides an estimation of the transmitted codeword. 

\section{Intelligent detectors}\label{SS:Train}

This section focuses on presenting intelligent detectors. Specifically, Section~\ref{SS:BP} presents a BP-based detector, while Section~\ref{SS:GNN} reports a graph neural network-based detector.  

\subsection{Belief propagation}\label{SS:BP}
Let $G_{P}=\left(\mathcal{V}\cup \mathcal{C}, \mathcal{E}\right)$ with $\mathcal{V}$, $\mathcal{C}$ and $\mathcal{E}$ respectively stand for the variable nodes (VNs), check nodes (CNs), edge nodes (ENs) of the Tanner graph $G_{P}$. Note that each row of $\mathbf{P}$ stands for a CN and each column for a VN. 
The Tanner graph can be seen as a deep neural network, in which the input layer receives the LLRs. The nodes in the hidden layers represent processing nodes. Each processing node is connected with a number of edges of the Tanner graph. As a consequence, each hidden layer consists of $E$ nodes. The number of nodes is equal to the size of $\mathcal{E}$. The output layer has $N$ processing elements and its responsibility is to provide an estimation of the transmitted~codeword.   

If the number of iterations is set to $L$, then the number of hidden layer are $2L$. The processing element $p_k$ of the hidden layer $k$ is associated with the VN, $v_k$, and CN, $c_k$, outputs~\cite{richardson_urbanke_2008}
\begin{align}
    t_{k,e_k} \hspace{-0.1cm}= \left\{
        \begin{array}{l l}
        l_{v_k} + \sum_{e^{'}_{k-1}, c_{k-1}\neq c_k} t_{k-1,e^{'}}, & \text{for } k \text{ odd} \\
        2\,\tan^{-1}\left(\prod_{e^{''}, v_{k-1}\neq v_k} \tanh\left(\frac{t_{k-1,e^{''}}}{2}\right)\right), &  \text{for } k \text{ even} 
    \end{array}\right.
\end{align}
where $e^{'}=\left(v_k, c_{k-1}\right)$, $e^{''}=\left(v_{k-1}, c_{k}\right)$ and $l_{v_k}$ is the self LLR message of $v_k$. The $k-$th node of the output layer reports
\begin{align}
    o_k = l_{v_{2L}} + \sum_{e^{'}} t_{2L, e^{'}}. 
\end{align}

\subsection{Graph neural network}\label{SS:GNN}

Similar to the BP approach, we consider a Tanner graph $G_g=\left(\mathcal{V}_g\cup \mathcal{F}_g, \mathcal{E}_g\right)$, where $\mathcal{V}_g$ is the set of the VNs. Each VN stands for a specific element of $\mathbf{c}$. $\mathcal{F}_g$ represents the set of the CNs, and $\mathcal{E}_g$ is the set of ENs. If ${P}_{i,j}=1$, then $v_{g,i}$ is connected to  $f_{g,j}$, where $P_{i,j}$ is the $i,j$ element of $\mathbf{P}$, $v_{g,i}$ and $f_{g,j}$ are the $i$ and $j$ elements of the sets $\mathcal{V}_g$, and  $\mathcal{F}_g$, respectively. To denote the set of all VNs that are connected to the $v_{g,i}$, we use $\mathcal{V}_g\left(v_{g,i}\right)$. Similarly, the set of all the CNs that are associated with $f_{g,j}$ is represented by $\mathcal{F}_g\left(f_{g,j}\right)$. 

To train the graph neural network, we use a function for the update of the edge messages and another one that update the nodes. Let $m_{v,i,j}$ be the updated message from $v_i$ to $v_j$, then
\begin{align}
    \mathbf{m}_{v,i,j}=g^{m}\left(\left[\mathbf{w}_{v_{g,i}}||\mathbf{w}_{f_{g,j}} || \mathbf{g}_{m_{v_{g,i},f_{g,j}}} \right], \mathbf{a}_{m_{v,f}} \right),
\end{align}
where $\mathbf{w}_{v_{g,i}}$ and $\mathbf{w}_{f_{g,j}}$ are vectors computed by each node for the VN $v_{g,i}$ and the CN $f_{g,j}$, respectively. Moreover, $\mathbf{a}_{m_{v,f}}$ stands for the trainable parameters. Finally, $g^{m}\left(\cdot\right)$ represents the parametrized function.  Let $m_{u,j,i}$ be the updated message value from the CN $u_j$ to the VN $v_j$. Thus, it can be obtain as
\begin{align}
    \mathbf{m}_{f,i,j}=g^{n}\left(\left[\mathbf{w}_{f_{g,j}}||\mathbf{w}_{v_{g,i}} || \mathbf{g}_{m_{f_{g,j},v_{g,i}}} \right], \mathbf{a}_{m_{f,v}} \right),
\end{align}
To evaluate the updated value of $\mathbf{w}_{v_{g,i}}$, we apply
\begin{align}
    \mathbf{w}_{v_{g,i}}^{'} = g^{n}\left(\left[\mathbf{w}_{v_{g,i}}|| \oplus_{v_i\in\mathcal{V}_g} \mathbf{m}_{v,i,j} || \mathbf{g}_{v_{g,i}}\right] \mathbf{a}_{v}\right), 
\end{align}
where $\mathbf{a}_{v}$ stands for the trainable parameters of the VN. Following a similar approach, the CN values can be updated~as  
\begin{align}
    \mathbf{w}_{f_{g,j}}^{'} =g^{n}\left(\left[\mathbf{w}_{f_{g,i}}|| \oplus_{f_i\in\mathcal{C}_g} \mathbf{m}_{f,i,j} || \mathbf{g}_{f_{g,i}}\right] \mathbf{a}_{f}\right),
\end{align}
where $\mathbf{a}_{v}$ and $\mathbf{a}_{f}$ are the VN and CN parameters, respectively. 

The training process consists of two phases: i) initialization, and ii) iterative optimization. In this paper, we employ the Gloron uniform initalizer~\cite{GlorotB10} to find the initial values of the trainable parameters, and the Adam optimizer to find their (sub)optimal values. As a loss function, the binary cross-entropy is~applied.  

\section{Results \& Discussion}\label{S:Results}

This section  presents Monte Carlo simulations that reveal the effectiveness of the ML-based detection approaches in the mitigating the impact of IQI and benchmarks GNN against BP and conventional approaches. The following scenario is considered. A wireless system that operates in the $120\,\mathrm{GHz}$ band and employs  low-density parity check code (LDPC) with code-rate equal to $0.714$. The parity check code has $63\times 45$ size. The zero-padding adds $1$ bit if and only if the length of the codeword is odd. A quadrature phase shift keying modulator is used by the TX and the corresponding demodulator by the RX. Both the TX and RX suffer from IQI with phase error equal to $5^o$. The BF and GNN respectively perform $20$ and $8$ iterations.  

\begin{figure}
	\centering
	\scalebox{0.66}{\begin{tikzpicture}
			\begin{axis}[
				ymode = log,
				xlabel={SNR (dB)},
				ylabel={BER},
				legend pos=south west,
				legend style={at={(1.4,0.75)},anchor=north,legend cell align=left},
				xmin = 2,
				xmax =9,
				ymin = 1e-5,
				ymax= 1e-1,
				ymajorgrids=true,
				xmajorgrids=true,
				grid style=dashed,
				]
				
				\addplot[
				color=black,
				mark=square]
				coordinates {
					(2.0, 2.2822e-02 )
					(3.0, 1.3737e-02 )
					(4.0, 7.4524e-03 )
					(5.0,  3.4825e-03 )
					(6.0, 1.4354e-03)
					(7.0, 4.7324e-04)
					(8.0, 1.2802e-04)
					(9.0, 2.5714e-05)
				};
				\addlegendentry{\small{Conventional (Ideal RF)}}
				
				\addplot[
				color=black,
				mark=o]
				coordinates {
					(2.0, 6.0303e-02)
					(3.0, 3.2900e-02)
					(4.0, 1.3673e-02)
					(5.0, 4.4496e-03)
					(6.0, 1.0363e-03)
					(7.0, 1.5273e-04)
					(8.0, 1.9825e-05)
					(9.0, 7.6190e-07)
				}  ;
				\addlegendentry{\small{BP (Ideal RF)}}

				\addplot[
				color=black,
				mark=star]
				coordinates {
					(2.0, 5.7614e-02)
					(3.0, 2.9835e-02)
					(4.0, 1.0817e-02)
					(5.0, 2.2960e-03)
					(6.0, 2.8305e-04)
					(7.0, 2.1556e-05)
					(8.0, 1.7302e-06)
					(9.0, 2.8571e-07)
				}  ;
				\addlegendentry{\small{GNN (Ideal RF)}}
				
				\addplot[
				color=red,
				mark=square]
				coordinates {
					(2.0, 5.1992e-02)
					(3.0, 3.3473e-02)
					(4.0, 1.7799e-02)
					(5.0, 7.7302e-03)
					(6.0, 3.5729e-03)
					(7.0, 1.1710e-03)
					(8.0, 2.6357e-04)
					(9.0, 4.5812e-05 )
				} ;
				\addlegendentry{\small{Conventional ($I_{t}=I_{r}=20\,\rm{dB}$)}}
				
				\addplot[
				color=red,
				mark=o]
				coordinates {
					(2.0, 2.4667e-02)
					(3.0, 1.0791e-02)
					(4.0, 3.9036e-03)
					(5.0, 9.7439e-04)
					(6.0, 2.2875e-04)
					(7.0, 4.6476e-05)
					(8.0, 1.0841e-05)
					(9.0, 1.3968e-06 )
				} ;
				\addlegendentry{\small{BP ($I_{t}=I_{r}=20\,\rm{dB}$)}}
				
				\addplot[
				color=red,
				mark=star]
				coordinates {
					(2.0, 2.2032e-02)
					(3.0, 8.3810e-03)
					(4.0, 1.9667e-03)
					(5.0, 3.2125e-04)
					(6.0, 3.5619e-05)
					(7.0, 3.5238e-06)
					(8.0, 3.6508e-07)
					(9.0, 7.9365e-08)
				}  ;
				\addlegendentry{\small{GNN ($I_{t}=I_{r}=20\,\rm{dB}$)} }
				
				\addplot[
				color=blue,
				mark=square]
				coordinates {
					(2.0, 3.7759e-02)
					(3.0, 2.3075e-02)
					(4.0, 1.2503e-02)
					(5.0, 5.9524e-03)
					(6.0, 2.3754e-03)
					(7.0, 7.8762e-04)
					(8.0, 1.8880e-04)
					(9.0, 3.2163e-05)
				} ;
				\addlegendentry{\small{Conventional ($I_{t}=I_{r}=30\,\rm{dB}$)}}
				
				\addplot[
				color=blue,
				mark=o]
				coordinates {
					(2.0, 6.0303e-02)
					(3.0, 3.2900e-02)
					(4.0, 1.3673e-02)
					(5.0, 4.4496e-03)
					(6.0, 1.0363e-03)
					(7.0, 1.5273e-04)
					(8.0, 1.9825e-05)
					(9.0, 7.6190e-07)
				} ;
				\addlegendentry{\small{BP ($I_{t}=I_{r}=30\,\rm{dB}$)}}
				
				\addplot[
				color=blue,
				mark=star]
				coordinates {
					(2.0, 6.5973e-02)
					(3.0, 4.2553e-02)
					(4.0, 1.7486e-02)
					(5.0, 3.4468e-03)
					(6.0, 3.6190e-04)
					(7.0, 5.1532e-05)
					(8.0, 3.1381e-06)
					(9.0, 4.0952e-07)
				} ;
				\addlegendentry{\small{GNN ($I_{t}=I_{r}=30\,\rm{dB}$)}}
				
			\end{axis}
			
	\end{tikzpicture}}
	\caption{BER vs SNR for different coding schemes and levels of IQI.}
	\label{Fig:BER}
\end{figure}
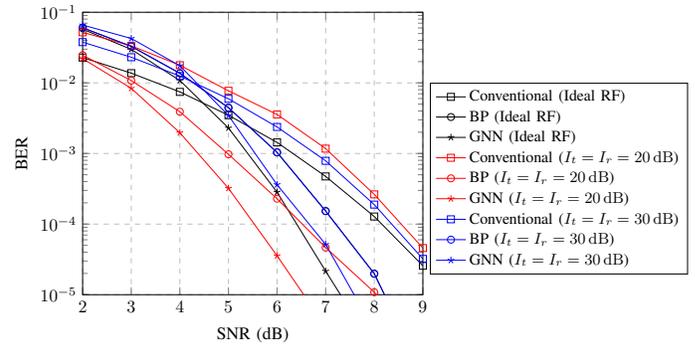

Figure~\ref{Fig:BER} depicts the BER as a function of the SNR for different IQI levels and coding/decoding schemes. As benchmarks, the cases of conventional detectors and ideal RF front-end are considered. As expected, for given detector and level of IQI, as the SNR increases, the error performance improves. For example, for the ideal RF front-end with conventional detector case, as the SNR increases from $7$ to $9\,\rm{dB}$, the BER decreases for more than one order of magnitude. For the same SNR variation and for the case in which a conventional detector is employed, but both the TX and RX suffer of IQI with IRR equal to $20\,\rm{dB}$, the BER decreases for about two orders of magnitude. Moreover, for conventional decoders and a SNR beyond $3\,\rm{dB}$, as the level of IQI increases, i.e., as IRR decreases, the error performance degrades. For instance, for a SNR equal to $7\,\rm{dB}$\,\rm{dB}, as IRR increases from $20$ to $30\,\rm{dB}$, the BER decreases from $1.17\times 10^{-3}$ to $7.88\times 10^{-4}$. On the other hand, for either BF or GNN-based decoders and a fixed SNR, as the level of IQI increases, an error performance improvement is observed. For example, for BF-based detector and SNR equals $7\,\rm{dB}$, the BER decreases from $1.53\times 10^{-4}$ to $4.45\times 10^{-5}$, as the IRR decreases from $30$ to $20\,\rm{dB}$. For the same SNR, but GNN-based detector, the BER decreases from $5.15\times 10^{-5}$ to $3.52\times 10^{-6}$, as the IRR decreases from $30$ to $20\,\rm{dB}$. As explained in~\cite{A:IQSC}, this is due to the TX IQI induced diversity order that can be exploited by the intelligent detectors. Additionally, from this figure, we observe that for ideal RF and a given SNR that is beyond $4.5\,\rm{dB}$, GNN-based detectors outperforms both BF-based and conventional detectors. On the other hand, in wireless systems in which their transceiver suffer from IQI, for any given SNR, GNN-based detectors outperforms both BF-based and conventional detectors. For instance, for SNR equals $7\,\rm{dB}$ and both the IRR of the TX and and RX equal to $20\,\rm{dB}$, the GNN-based detector achieves a BER that is equal to $3.52\times 10^{-6}$, while, for the same SNR and IRR, BF-based detectors achieve a BER that equals $4.65\times 10^{-5}$, and conventional detectors achieve a BER, which is equal to $1.17\times 10^{-3}$. Notice that the GNN-based detectors uses $8$ iterations, while the BF-based one $20$. In other words, with less iterations the GNN-based detector achieves better performance than the BF-based detector.

\section{Conclusions}\label{S:Con}
In this paper, we presented a GNN-based intelligent detector and demonstrated its ability to prevent IQI. Specifically, we consider a two-way wireless system in which the TX and RX are both susceptible to IQI at high frequencies and in opposite directions. The transmitter uses a linear error correction code, while the receiver employs a GNN-based decoder. The BER is calculated via independent Monte Carlo simulations in order to quantify the system's performance. The results are compared against those wireless systems that rely on traditional and BF based detectors, and illustrate the performance improvements that can be achieved when employing the proposed GNN-based~detector.

\section*{Acknowledgement}
This work has received funding from the European Unions Horizon-CL4-2021 research and innovation programme under grant agreement No. 101070181 (TALON).

\balance
\bibliographystyle{IEEEtran}
\bibliography{IEEEabrv,References}

\begin{thebibliography}{10}
\providecommand{\url}[1]{#1}
\csname url@samestyle\endcsname
\providecommand{\newblock}{\relax}
\providecommand{\bibinfo}[2]{#2}
\providecommand{\BIBentrySTDinterwordspacing}{\spaceskip=0pt\relax}
\providecommand{\BIBentryALTinterwordstretchfactor}{4}
\providecommand{\BIBentryALTinterwordspacing}{\spaceskip=\fontdimen2\font plus
\BIBentryALTinterwordstretchfactor\fontdimen3\font minus
  \fontdimen4\font\relax}
\providecommand{\BIBforeignlanguage}[2]{{%
\expandafter\ifx\csname l@#1\endcsname\relax
\typeout{** WARNING: IEEEtran.bst: No hyphenation pattern has been}%
\typeout{** loaded for the language `#1'. Using the pattern for}%
\typeout{** the default language instead.}%
\else
\language=\csname l@#1\endcsname
\fi
#2}}
\providecommand{\BIBdecl}{\relax}
\BIBdecl

\bibitem{Boulogeorgos2018}
A.-A.~A. Boulogeorgos, A.~Alexiou, T.~Merkle, C.~Schubert, R.~Elschner,
  A.~Katsiotis, P.~Stavrianos, D.~Kritharidis, P.~K. Chartsias, J.~Kokkoniemi,
  M.~Juntti, J.~Lehtom\"aki, A.~Teixeir\'a, and F.~Rodrigues, ``Terahertz
  technologies to deliver optical network quality of experience in wireless
  systems beyond {5G},'' \emph{IEEE Commun. Mag.}, vol.~56, no.~6, pp.
  144--151, Jun. 2018.

\bibitem{A:Analytical_Performance_Assessment_of_THz_Wireless_Systems}
A.-A.~A. Boulogeorgos, E.~N. Papasotiriou, and A.~Alexiou, ``Analytical
  performance assessment of {THz} wireless systems,'' \emph{IEEE Access},
  vol.~7, no.~1, pp. 1--18, Jan. 2019.

\bibitem{Boulogeorgos2019}
A.-A.~A. Boulogeorgos and A.~Alexiou, ``Analytical performance evaluation of
  beamforming under transceivers hardware imperfections,'' in \emph{{IEEE}
  Wireless Communications and Networking Conference ({WCNC})}.\hskip 1em plus
  0.5em minus 0.4em\relax {IEEE}, Apr. 2019.

\bibitem{9615497}
T.~A. Tsiftsis, C.~Valagiannopoulos, H.~Liu, A.-A.~A. Boulogeorgos, and N.~I.
  Miridakis, ``Metasurface-coated devices: A new paradigm for energy-efficient
  and secure 6g communications,'' \emph{IEEE Vehicular Technology Magazine},
  vol.~17, no.~1, pp. 27--36, 2022.

\bibitem{trevlakis2023localization}
S.~E. Trevlakis, A.-A.~A. Boulogeorgos, D.~Pliatsios, K.~Ntontin,
  P.~Sarigiannidis, S.~Chatzinotas, and M.~D. Renzo, ``Localization as a key
  enabler of {6G} wireless systems: {A} comprehensive survey and an outlook,''
  2023.

\bibitem{A:Effects_of_RF_impairments_In_Cascaded}
A.-A.~A. Boulogeorgos, P.~C. Sofotasios, B.~Selim, S.~Muhaidat, G.~K.
  Karagiannidis, and M.~Valkama, ``Effects of {RF} impairments in
  communications over cascaded fading channels,'' \emph{{IEEE} Trans. Veh.
  Technol.}, vol.~65, no.~11, pp. 8878 -- 8894, Nov. 2016.

\bibitem{9039743}
E.~N. Papasotiriou, A.-A.~A. Boulogeorgos, and A.~Alexiou, ``Performance
  analysis of thz wireless systems in the presence of antenna misalignment and
  phase noise,'' \emph{IEEE Communications Letters}, vol.~24, no.~6, pp.
  1211--1215, 2020.

\bibitem{8094903}
A.-A.~A. Boulogeorgos and G.~K. Karagiannidis, ``Energy detection in
  full-duplex systems with residual rf impairments over fading channels,''
  \emph{IEEE Wireless Communications Letters}, vol.~7, no.~2, pp. 246--249,
  2018.

\bibitem{7460238}
A.-A.~A. Boulogeorgos, D.~S. Karas, and G.~K. Karagiannidis, ``How much does
  i/q imbalance affect secrecy capacity?'' \emph{IEEE Communications Letters},
  vol.~20, no.~7, pp. 1305--1308, 2016.

\bibitem{9159653}
A.~A. Boulogeorgos and A.~Alexiou, ``How much do hardware imperfections affect
  the performance of reconfigurable intelligent surface-assisted systems?''
  \emph{IEEE Open Journal of the Communications Society}, vol.~1, pp.
  1185--1195, 2020.

\bibitem{7496991}
A.-A.~A. Boulogeorgos, H.~A.~B. Salameh, and G.~K. Karagiannidis, ``Spectrum
  sensing in full-duplex cognitive radio networks under hardware
  imperfections,'' \emph{IEEE Transactions on Vehicular Technology}, vol.~66,
  no.~3, pp. 2072--2084, 2017.

\bibitem{9145251}
J.~Kokkoniemi, A.-A.~A. Boulogeorgos, M.~U. Aminu, J.~Lehtomäki, A.~Alexiou,
  and M.~Juntti, ``Stochastic analysis of indoor thz uplink with co-channel
  interference and phase noise,'' in \emph{2020 IEEE International Conference
  on Communications Workshops (ICC Workshops)}, 2020, pp. 1--6.

\bibitem{8933807}
B.~Batagelj, J.~Capmany, and E.~G. Udvary, ``5th-generation mobile access
  networks assisted by integrated microwave photonics,'' in \emph{2019
  International Workshop on Fiber Optics in Access Networks (FOAN)}, 2019, pp.
  1--6.

\bibitem{A:Frequency_selective_IQ_mismatch_callibration_of_wideband_DC_transmitters}
L.~Anttila, M.~Valkama, and M.~Renfors, ``Frequency-selective {I/Q} mismatch
  calibration of wideband direct-conversion transmitters,'' \emph{IEEE Trans.
  Circuits Syst. II Express Briefs}, vol.~55, no.~4, pp. 359--363, Apr. 2008.

\bibitem{8527635}
M.~Aziz, M.~Vejdani~Amiri, M.~Helaoui, and F.~M. Ghannouchi, ``Statistics-based
  approach for blind post-compensation of modulator’s imperfections and power
  amplifier nonlinearity,'' \emph{IEEE Transactions on Circuits and Systems I:
  Regular Papers}, vol.~66, no.~3, pp. 1063--1075, 2019.

\bibitem{8383719}
D.~Wang, M.~Aziz, M.~Helaoui, and F.~M. Ghannouchi, ``Augmented real-valued
  time-delay neural network for compensation of distortions and impairments in
  wireless transmitters,'' \emph{IEEE Transactions on Neural Networks and
  Learning Systems}, vol.~30, no.~1, pp. 242--254, 2019.

\bibitem{9605580}
Y.~Wu, U.~Gustavsson, A.~G.~I. Amat, and H.~Wymeersch, ``Low complexity joint
  impairment mitigation of i/q modulator and pa using neural networks,''
  \emph{IEEE Journal on Selected Areas in Communications}, vol.~40, no.~1, pp.
  54--64, 2022.

\bibitem{8476222}
P.~Jaraut, M.~Rawat, and F.~M. Ghannouchi, ``Composite neural network digital
  predistortion model for joint mitigation of crosstalk, $i/q$ imbalance,
  nonlinearity in mimo transmitters,'' \emph{IEEE Transactions on Microwave
  Theory and Techniques}, vol.~66, no.~11, pp. 5011--5020, 2018.

\bibitem{7463533}
A.-A.~A. Boulogeorgos, N.~D. Chatzidiamantis, and G.~K. Karagiannidis, ``Energy
  detection spectrum sensing under rf imperfections,'' \emph{IEEE Transactions
  on Communications}, vol.~64, no.~7, pp. 2754--2766, 2016.

\bibitem{B:Schenk-book}
T.~Schenk, \emph{{RF} Imperfections in High-Rate Wireless Systems}.\hskip 1em
  plus 0.5em minus 0.4em\relax The Netherlands: Springer, 2008.

\bibitem{PhD:Boulogeorgos}
A.-A.~A. Boulogeorgos, ``Interference mitigation techniques in modern wireless
  communication systems,'' Ph.D. dissertation, Aristotle University of
  Thessaloniki, Thessaloniki, Greece, Sep. 2016.

\bibitem{8877185}
A.-A.~A. Boulogeorgos and A.~Alexiou, ``Performance evaluation of the initial
  access procedure in wireless thz systems,'' in \emph{2019 16th International
  Symposium on Wireless Communication Systems (ISWCS)}, 2019, pp. 422--426.

\bibitem{richardson_urbanke_2008}
T.~Richardson and R.~Urbanke, \emph{Modern Coding Theory}.\hskip 1em plus 0.5em
  minus 0.4em\relax Cambridge University Press, 2008.

\bibitem{GlorotB10}
\BIBentryALTinterwordspacing
X.~Glorot and Y.~Bengio, ``Understanding the difficulty of training deep
  feedforward neural networks.'' in \emph{AISTATS}, ser. JMLR Proceedings,
  Y.~W. Teh and D.~M. Titterington, Eds., vol.~9.\hskip 1em plus 0.5em minus
  0.4em\relax JMLR.org, 2010, pp. 249--256. [Online]. Available:
  \url{http://dblp.uni-trier.de/db/journals/jmlr/jmlrp9.html}
\BIBentrySTDinterwordspacing

\bibitem{A:IQSC}
A.-A.~A. Boulogeorgos, V.~M. Kapinas, R.~Schober, and G.~K. Karagiannidis,
  ``{I/Q}-imbalance self-interference coordination,'' \emph{IEEE Trans.
  Wireless Commun.}, vol.~15, no.~6, pp. 4157 -- 4170, Jun. 2016.

\end{thebibliography}

\end{document}